\documentclass[10pt,a4paper,twoside]{article}
\usepackage{epsfig}
\usepackage{baltlat6}
\usepackage{array}
\def\aap{\mbox{A\&A}}

\begin{document}
\ \
\vspace{0.5mm}
\setcounter{page}{277}
\vspace{8mm}

\titlehead{Baltic Astronomy, vol.\,20, 130--134, 2011}

\titleb{AGN DUSTY TORI AS A CLUMPY TWO-PHASE MEDIUM:\\THE 10 $\mu$m
SILICATE FEATURE}

\begin{authorl}
\authorb{Marko Stalevski}{1,2}
\authorb{Jacopo Fritz}{2}
\authorb{Maarten Baes}{2}
\authorb{Theodoros Nakos}{2} and
\authorb{Luka \v{C}. Popovi\'{c}}{1}
\end{authorl}

\begin{addressl}
\addressb{1}{Astronomical Observatory,\\ Volgina 7, 11060 Belgrade,
Serbia; mstalevski@aob.rs}
\addressb{2}{Sterrenkundig Observatorium, Universiteit Gent,\\
Krijgslaan 281-S9, Gent, 9000, Belgium}
\end{addressl}

\submitb{Received: 2011; accepted: 2011}

\begin{summary}
We investigated the emission of active galactic nuclei dusty
tori in the infrared domain, with a focus on the $10$ $\mu$m
silicate feature. We modeled the dusty torus as a clumpy
two-phase medium with high-density clumps and a low-density medium
filling the space between the clumps. We employed a three-dimensional
radiative transfer code to obtain spectral energy distributions and
images of tori at different wavelengths. We calculated a grid of
models for different parameters and analyzed the influence of these
parameters on the shape of the mid-infrared emission. A corresponding
set of clumps-only models and models with a smooth dust distribution
is calculated for comparison. We found that the dust distribution,
the optical depth and a random arrangement of clumps in the innermost
region, all have an impact on the shape and strength of the silicate
feature. The $10$ $\mu$m silicate feature can be suppressed for some
parameters, but models with smooth dust distribution are also able to
produce a wide range of the silicate feature strength.
\end{summary}


\resthead{AGN dusty tori as a clumpy two-phase medium: the 10 $\mu$m
silicate feature}
{M. Stalevski et al.}

\sectionb{1}{INTRODUCTION}

The dusty torus surrounding the central engine of an active galactic
nucles (AGN), absorbs the incoming accretion disc radiation and
re-emits it in the infrared domain. As a result, a mid- to
far-infrared bump is observed in the spectral energy distribution
(SED) of AGNs and a silicate feature caused by Si-O stretching modes,
giving rise to either an emission or absorption feature, peaking at
$\sim 10$ $\mu$m. In type 1 sources, hot dust in the inner region can
be seen directly and the feature is expected to be detected in
emission. Mid-infrared observations obtained with the
\textit{Spitzer} satellite confirm the silicate emission feature in
AGNs (e.g. Siebenmorgen et al. 2005; Hao et al. 2005). In type 2
objects, the silicate feature is usually observed in absorption
(e.g., Jaffe et al. 2004) due to obscuration by the cold dust.

In order to prevent the dust grains from being destroyed by the hot
surrounding gas, Krolik \& Begelman (1988) suggested that the dust in
the torus is organized in a large number of optically thick clumps.
Wada \& Norman (2002) (with a model update in Wada et al. 2009)
performed a 3D hydrodynamical simulations of AGN tori, taking into
account the self-gravity of the gas, the radiative cooling, and
the heating due to supernovae. They found that such a turbulent
medium would produce a multi-phase filamentary (sponge-like)
structure, rather then isolated clumps.

We took a step further toward a more realistic model by treating the
dusty torus as a two-phase medium, with high density clumps and a low
density medium filling the space between them. We calculated SEDs and
images of the torus for a grid of parameters. Our approach allows us
to, for each two-phase model, to generate a clumps-only model (with
dust distributed in the clumps exclusively, without any dust between
them) and a smooth model with the same global physical parameters.
Our aims are (a)\, to investigate the influence of the different
parameters on model SEDs and their observable properties, with a
focus on the $10$ $\mu$m silicate feature, (b)\, to put to a test
claims that the observed SEDs in the mid-infrared domain
unambiguously point to a clumpy structure of dusty tori (e.g. Horst
et al. 2006); if that is indeed the case, a comparison of clumpy and
smooth models should show a systematic difference of their observable
properties, such as the strength of the silicate feature.

\sectionb{2}{MODEL}

\begin{figure}
\centering
\includegraphics[height=0.53\textwidth]{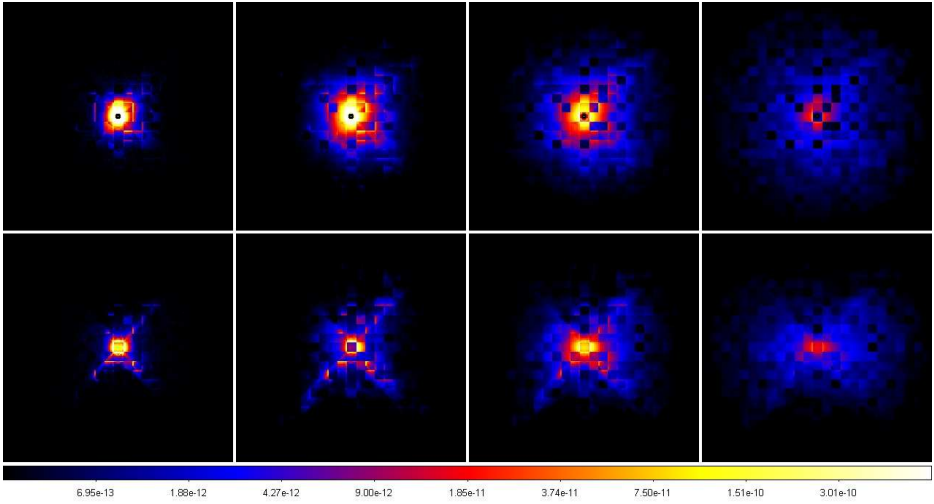}
\caption{Images of the torus at different wavelengths. Top row is
face-on view, bottom row edge-on view. From left to right, panels
represent images at $4.6$, $9.7$, $13.9$, and $30.7$ $\mu$m. Images
are given in logarithmic color scale. The visible squared structure
is due to the clumps which in the model are in the form of cubes.}
\label{fig:img}
\end{figure}

The description of the model an its parameters is given in Stalevski
et al (2011); here we will present only the general approach. We have
used the radiative transfer code SKIRT (Baes et al. 2003, 2011) for
the modelling of AGN dusty tori. SKIRT is a 3D continuum radiative
transfer code based on the Monte Carlo algorithm. As hydrodynamical
simulations of Wada \& Norman (2002) demonstrated, dust in tori is
expected to take the form of a multi-phase structure, rather than
isolated clumps. To generate dusty tori with a two-phase structure,
we start from smooth models and apply the algorithm described by Witt
\& Gordon (1996). According to this algorithm, each individual cell
in the grid is assigned randomly to either a high- or low-density
state by a Monte Carlo process. The medium created in such a way is
statistically homogeneous, but clumpy. This leads to the appearance
of complex structures formed by several merged clumps, clusters of
clumps, or even a single, interconnected sponge-like structure. To
tune the density of the clumps and the inter-clump medium, we use the
`contrast parameter', defined as the ratio of the dust density in the
high- and low-density phase. Setting extremely high value of contrast
($>1000$) effectively puts all the dust into the clumps, without any
low-density medium between them. We generated three sets of models
with the same global physical parameters: (a)\,models with the dust
distributed smoothly, (b)\,models with the dust as a two-phase medium
and (c)\, models with a contrast parameter set to an extremely high
value ($10^9$), effectively putting all the dust into the
high-density clumps. We will refer to these models as `smooth',
`two-phase' and `clumps-only', respectively.

\begin{figure}
\centering
\includegraphics[height=0.54\textwidth]{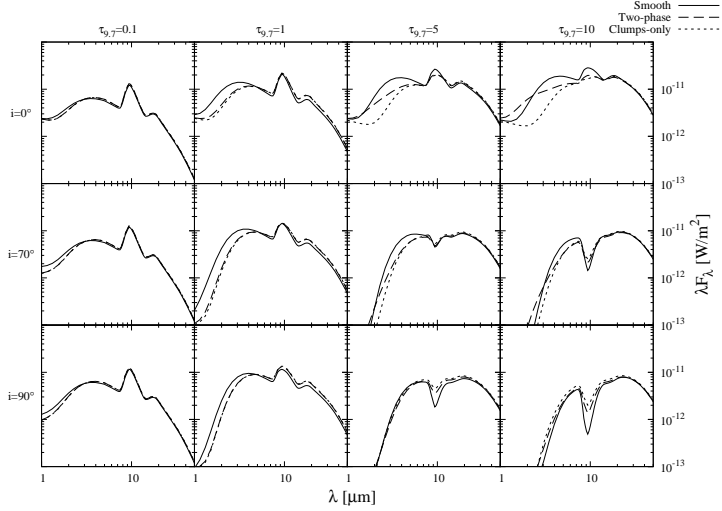}
\caption{Example of model SEDs in the $1-50$ $\mu$m wavelength range.
Solid line: smooth models; dashed line: two-phase models; dotted
line: clumps-only models. The columns correspond to optical depths of
$\tau_{9.7}=0.1,1,5,10.0$, from left to right. The rows correspond to
inclinations of $i=0,70,90^\circ$, from top to bottom. The dust
distribution parameters $p=1$ and $q=2$.}
\label{fig:sed}
\end{figure}

\newpage
\sectionb{3}{RESULTS AND CONCLUSIONS}

We calculated a grid of models\footnote{Model SEDs, in the form
of \texttt{ascii} files
are available on the following address:
\texttt{https://sites.google.com/site/skirtorus/}. Images, in the
form of \texttt{fits} files are available upon request.} for
different parameters, determining the total amount of dust, dust
distribution, torus size, clump size and random distribution of
clumps. Each model is calculated at inclinations of $0^\circ$,
$40^\circ$, $50^\circ$, $60^\circ$, $70^\circ$, $80^\circ$ and
$90^\circ$, where $i=0^\circ$ represents a face-on (type 1) AGN and
$i=90^\circ$ an edge-on (type 2) AGN. In order to analyze the
properties of the SEDs and the $10$ $\mu$m silicate feature, for each
model a number of parameters is measured, such as, the strength of
the silicate feature, the SED width, the isotropy of infrared
emission and the peak of the infrared emission. Fig. 1 shows images
of the torus at different wavelenghts, for face-on and edge-on
orientations. An example of SEDs for different values of optical
depth, inclination and for the three dust
configurations is presented in Fig 2. From the analysis of the SED
properties and the comparison of the corresponding models, we
conclude the following (Stalevski et al, 2011):

\begin{itemize}

\item The SED at near- and mid-infrared wavelengths is determined
by the conditions of the dust in the innermost region of the torus:
different random distributions of the clumps may result in the very
different SEDs in otherwise identical models.

\item The clump size has a major impact on the SED properties. SEDs
of the clumpy models with a large number of small clumps are very
similar to the ones obtained by a smooth dust distribution. In clumpy
models with bigger clumps, the silicate feature in absorption is less
deep; when present in emission, the silicate feature is generally
less pronounced. The suppression strongly depends on the dust
distribution parameters.

\item The shape of the silicate feature is not only a function of
inclination. Optical depth, dust distribution parameters, clump size
and actual arrangement of the clumps, all have an impact on the
appearance of the silicate feature. Low optical depth tori produce
silicate feature in a strong emission. Models with high-density
clumps occupying the innermost region will have the emission feature
attenuated due to the shadowing effects.

\item Although the silicate emission feature could be suppressed in
the clumpy models for certain parameters, the smooth models are able
to reproduce almost the same range of the silicate feature strength.
Our analysis shows that, overall, when considering characteristics of
the silicate feature, models with the three dust configurations
(smooth, two-phase, clumps-only) are not distinguishable.

\thanks{This work was supported by the European Commission (Erasmus
Mundus Action 2 partnership between the European Union and the
Western Balkans, http://www.basileus.ugent.be) and by the Ministry of
Education and Science of Serbia through the projects `Astrophysical
Spectroscopy of Extragalactic Objects' (146001) and `Gravitation and
the Large Scale Structure of the Universe' (146003).}

\end{itemize}

\References

\refb Baes M., et al., 2003, MNRAS, 343, 1081

\refb Baes M., Verstappen J., De Looze I., Fritz J., Saftly W., Vidal
P\'erez E., Stalevski M., Valcke S., 2011, ApJS, 196, 22

\refb Hao L., et al., 2005, ApJL, 625, L75

\refb Horst H., Smette A., Gandhi P., Duschl W. J., 2006, \aap, 457,
L17

\refb Jaffe W., et al., 2004, Nature, 429, 47

\refb Krolik J. H., Begelman M. C., 1988, ApJ, 329, 702

\refb Siebenmorgen R., Haas M., Kr{\"u}gel E., Schulz B., 2005, \aap,
436, L5

\refb Stalevski M., Fritz J., Baes M., Nakos T., Popovi\'{c}
L. \v{C}., 2011, MNRAS, accepted = arXiv:1109.1286

\refb Wada K., Norman C. A., 2002, ApJL, 566, L21

\refb Wada K., Papadopoulos P. P., Spaans M., 2009, ApJ, 702, 63

\refb Witt A. N., Gordon K. D., 1996, ApJ, 463, 681

\end{document}